\documentclass[fleqn,usenatbib]{mnras}

\usepackage{newtxtext,newtxmath}
\usepackage{subcaption}
\usepackage[T1]{fontenc}

\DeclareRobustCommand{\VAN}[3]{#2}
\let\VANthebibliography\thebibliography
\def\thebibliography{\DeclareRobustCommand{\VAN}[3]{##3}\VANthebibliography}

\usepackage{graphicx}	
\usepackage{amsmath}	

\newcommand{\kms}{km\,s$^{-1}$}

\title[Spectroscopy of M104 Satellite Candidates]{IFUM Integrated Field Spectroscopy of Ten M104 Satellite Galaxy Candidates}

\author[E.~Crosby et al.]{Ethan Crosby$^{1}$\thanks{E-mail: Ethan.Crosby@anu.edu.au},Mario Mateo$^{2}$,
Ivanna Escala$^{3}$,
Helmut Jerjen$^{1}$,
Oliver M\"uller$^{4}$,
Marcel S. Pawlowski$^{5}$
\\
$^{1}$Research School of Astronomy and Astrophysics, Australian National University, Canberra, ACT 2611, Australia\\
$^{2}$Department of Astronomy, University of Michigan, 1085 S. University Ave., Ann Arbor, MI 48109, USA \\
$^{3}$Department of Astrophysical Sciences, Princeton University, 4 Ivy Lane, Princeton, NJ 08544, USA \\
$^{4}$Institute of Physics, Laboratory of Astrophysics, Ecole Polytechnique F\'ed\'erale de Lausanne (EPFL), 1290 Sauverny, Switzerland\\
$^{5}$Leibniz-Institut f\"ur Astrophysik Potsdam, An der Sternwarte 16, D-14482 Potsdam, Germany\\
}
\date{Accepted XXX. Received YYY; in original form ZZZ}

\pubyear{2024}

\begin{document}
\label{firstpage}
\pagerange{\pageref{firstpage}--\pageref{lastpage}}
\maketitle

\begin{abstract}
We report the spectroscopic analysis of ten satellite galaxy candidates in the sphere of influence of the Sombrero galaxy (M104, NGC4594), based on data obtained with IFUM (Integral Field Units for Magellan). Based on their newly-observed recessional velocities, we confirm that nine of these candidates are satellite galaxies of M104, with one being a background dwarf galaxy. All ten dwarfs have stellar masses $2\times10^{7}\,M_{\odot}$ to $1\times10^{9}\,M_{\odot}$ and mean weighted metallicities $-1.7<\langle{[\mathrm{M/H}]}\rangle<-0.3$. Although these dwarfs are predominantly old, with stellar populations $\sim5-11\,$Gyr. However, this sample contains a local example of a low-mass "Green Pea" candidate, it exhibits extreme optical emission features and broad emission line features ($\sigma\sim250\,\mathrm{km\,s^{-1}}$) reminiscent of high-redshift Ly$\alpha$/LyC photon leaking galaxies. Using the newly-acquired recessional velocities of the nine satellites of M104, we find no evidence of coherent satellite motions unlike other nearby $L_*$ galaxy environments. Given the small sample, this results does not statistically rule out such coherent motions. There remain 60 satellite candidates of M104 for which future spectroscopy can more reliably test for such motion. Using the observed dwarf galaxies as tracers of the gravitational potential of M104, we estimate the dynamical mass of M104, $M_{dyn}=(12.4\pm6.5)\times10^{12}M_{\odot}$, and find that, making a reasonable estimate of M104's gas mass, $>90$\% of its baryons are missing. These results agree with previous measurements of M104's dynamical mass.

\end{abstract}



\section{Introduction}
Reconstruction of the satellite galaxy systems of the Milky Way and M31 with full 6D phase-space coordinates consisting of the on-sky positions, distances, line-of-sight (LOS) velocities and on-sky proper motions has revealed that these satellites are co-orbiting about their hosts in flattened, apparently stable structures known as satellite planes \citep{Ibata_2013,Conn2013,Pawlowski2015,Pawlowski_2019,Pawlowski_2021_b}. Based on only positions and radial velocities, a similar configuration of satellites appears to exist around Centaurus A \citep{Tully2015, Mueller2018, Mueller_2019b, Mueller2021}, NGC253 \citep{MartnezDelgado2021} and NGC4490/NGC4485 \citep{Karachentsev2024}. These empirical results for well-studied host galaxies in the local universe suggest that satellite planes may be common.  In contrast, similarly coherent, flattened, and (apparently) stable structures are rarely seen in Lambda Cold Dark Matter ($\Lambda$CDM) cosmological simulations which generally predict that satellites do not often have preferential spatial and kinematic orientations \citep{Metz_2009_a,Ibata_2013,Ibata2014,Mueller2018,Pawlowski_2019,Pawlowski_2021_a,Pawlowski_2021_b}. 

Extensive research has been conducted to better understand the discrepancy. \cite{Cautun2015} highlighted statistical biases in the methods used to determine the significance of a plane, specifically the \textit{look-elsewhere} effect. This occurs when apparently significant results are found purely by chance when evaluating a large parameter space, which may have produced a misleading level of significance in the evaluation of the Local Group satellite planes.  Other conceivable biases exist that could explain coherent satellite structures. For example, one may argue unique formation conditions within the local cosmic environment (e.g. the Local Sheet: \citealt{Tully2008}) that hosts the best-studied galaxies with satellite planes to date favour the formation of coherent structures around nearby host galaxies. \cite{Libeskind_2019} found a preferential global alignment of satellites within the Local Sheet, and other studies \citep{Neuzil2020,AragonCalvo2022} are finding that the structure of the Local Sheet may be a cosmological outlier. These solutions mostly pertain to a class of physical processes, including galaxy accretion along flattened cosmic filaments \citep{Libeskind2010,Lovell2011}, group infall of dwarf galaxies \citep{Metz2009_B,Vasiliev2023} and interaction fragments, known as tidal dwarf galaxies, rotating in thin planes following galactic interactions \citep{Kroupa_2005,Metz2007_b,Pawlowski2012,Kroupa_2012,Hammer_2013,Blek2018,Banik2022}. Other potential solutions suggest that observed satellite planes are transient or unstable structures, based on searches for satellite planes within simulations \citep{Bahl_2014,Gillet_2015,Buck_2016,Shao_2019}. There are also a growing number of studies that suggest the satellite plane of the Milky Way is not a outlier in $\Lambda$CDM simulations when the dynamics of the Large Magellanic Cloud are considered \citep{Samuel2021,GaravitoCamargo2021,Vasiliev2023}, though this specific explanation obviously does not directly apply to other systems where coherent dwarf structures are claimed to exist.

Questions that arise from this summary include: Which of these solutions represent 'normal' dwarf galaxy distributions?  Does one specific bias or physical effect dominate as the source of the model/observation discrepancy, or are various factors simultaneously involved?  Addressing these questions requires a broader range of observed systems that are located in a larger, more diverse region of space beyond the local volume.  Aside from some recent pioneering efforts \citep{Heesters2021, Karachentsev2024}, based on spatial distributions, few comprehensive studies of satellite systems beyond a few {\it Mpc} have been carried out to date.  This has motivated searches for dwarf galaxy systems associated with hosts located well beyond the Local Volume and, so far, out to distances of up to $12\ \textsc{Mpc}$ \citep{Javanmardi_2016,Crnojevi2016,Muller_2017,Muller_2018_c,Danieli2020,Byun_2020,Muller_2020,Carlsten_2022,Crosby_2023_a,Crosby_2023_b}.  Extensive catalogs of dwarf satellite galaxy candidates out to this distance now exist as a result of these wide and deep optical surveys.  But that is only the first step.  Determining the presence or absence of a satellite plane requires full positional (membership) and kinematic information for these satellites. 

The next crucial step to explore the frequency and general properties of satellite distributions--including the key question of whether satellite planes are ubiquitous--is to carry out spectroscopic follow-up of satellite candidates \citep{Muller2021, Heesters2023}. Spectroscopy serves three key purposes: (i) confirmation of candidates as true satellites of their assumed hosts; (ii) exploring the satellite kinematics about their hosts; (iii) discerning the metallicity, stellar content, and—in favorable cases, internal kinematics of satellite systems. These results help to distinguish true satellites from background galaxies and to look for possible co-rotation of satellites about their host, a key signature of satellite planes. Ultimately, such studies of newly-discovered satellite systems also broaden the scope of interesting research questions, including determining how the morphology of the host galaxy may affect the prevalence of a satellite plane, whether satellite-plane membership of dwarfs correlates to star-formation history (as expected in hierarchical formation), how dwarfs are internally affected by group membership, and whether larger-scale cosmic superstructures influence the formation of satellite planes \citep{Pawlowski2018,Libeskind_2019}.

In this paper, we present the initial results of an observational study of dwarf satellite candidates in the M104 group aimed at addressing these science goals. In section 2 we introduce ten dwarf satellite galaxy candidates identified photometrically near M104 and describe the use of IFUM (Integral Field Units for Magellan) to study these systems spectroscopically. In Section 3 we present results of the observations on these ten candidates with IFUM, including an investigation of whether this sample reveals coherent satellite motions in the halo of M104, and a determination of the dynamical mass of M104 based on the kinematics of its dwarf galaxy population.

\section{Target Selection}

In a previous study \citep{Crosby_2023_b}, we searched for satellite galaxies within the virial radius of the Sombrero galaxy (M104, also known as NGC4594 or PGC42407) using deep CCD $g$-band images obtain with the Hyper Suprime-Cam (HSC) at the 8m Subaru telescope. Combined with other imaging surveys conducted around M104 aimed at finding satellite galaxy candidates \citep{Karachentsev_2000,Javanmardi_2016,Carlsten_2020,Karachentsev2020_b,Crosby_2023_b} we can identify 75 dwarf satellite galaxy candidates within the virial radius of M104, complete to a magnitude limit of $M_g\sim-9.5$ or a surface brightness limit of $\mu_{0,g}\sim27$. 

This sample is ideal for follow-up spectroscopic or resolved and deep photometric observations to confirm how many are likely physical satellites of M104. Given that M104 is largely isolated from other $L_*$ galaxies, systemic velocity measurements are not only a reliable means of confirming which of these dwarf galaxy candidates are associated with M104 but also provides a means of exploring whether any substructure--such as satellite planes--are present. The large size of the M104 sample has particular potential to greatly improve the statistical analysis of the system. Its isolation from other massive host galaxies creates an ideal setting to examine its satellite population to explore the satellite plane problem. These factors have motivated the spectroscopy follow-up in this paper.

M104 is a particularly well-suited for follow-up spectroscopy for a number of reasons. The galaxy resides at a reasonably close distance of $9.55\pm0.34\,\mathrm{Mpc}$ \citep{McQuinn_2016} and has a distinct systemic recessional velocity of v$_\odot=1095\pm5\,\mathrm{km\,s^{-1}}$. With a high peculiar velocity of $v_p\sim300-400\,\mathrm{km\,s^{-1}}$, this places this galaxy well outside the Local Sheet as delineated by a line-of-sight (LOS) recessional velocity discontinuity in galaxies beyond $\sim7-8\,\mathrm{Mpc}$ \citep{Tully2008,McCall2014,Karachentsev_2015_A,Anand2019}. Given the satellite planes associated with galaxies in the Local Sheet appear to align with the Local Sheet \citep{Libeskind_2019}, M104 represents an interesting test case that should not be affected by this local cosmological structure. M104 is also spatially and kinematically distinct from the Virgo Cluster given its location well in the foreground of the cluster’s southern extension by around $6-10\,\mathrm{Mpc}$ \citep{Tully_1982,Kourkchi_2017} and M104's considerably lower systemic velocity compared to the mean redshift of Virgo. This firmly places M104 in a sparsely-populated ’bridge’ between the Local Sheet and the core of the Virgo cluster. We list some of the basic characteristics of M104 in Table \ref{tab:M104param}.

Prior to this work, only five of the 75 current dwarf galaxy candidates have been spectroscopically confirmed as satellites within the M104 virial radius limit; these system's properties are summarized in Table 3. Of the remaining 70 objects, 39 are classified by \cite{Crosby_2023_b} as ’high’ probability M104 group members while the remaining 31 are considered to be ’low’ probability members. 

This paper presents new spectroscopic results for ten of these high-probability candidate group members (these are illustrated in Figure \ref{fig:M104_spaxel_coadds}). The data presented here were obtained with IFUM (Integral Field Units for Magellan), a new instrument deployed at the NasmythEast focus of the 6.5m Magellan/Clay telescope at Las Campanas Observatory \citep{Mateo2022}. These ten targets were selected to reasonably represent the diversity of dwarf galaxies found near M104, from vigorously star forming Blue Compact Dwarfs (BCD), to quenched low surface brightness Dwarf Ellipticals (dE). This variety of astrophysically interesting galaxies were also chosen, in part, to help evaluate the performance of IFUM in observing faint, low-surface brightness systems. The sample was also chosen to ensure a relatively even distribution of satellites around M104 as an exploration of the kinematics of the dwarf-galaxy system surrounding M104, and to serve as a first attempt to detect the presence of a possible satellite plane, or other substructure, around M104.

\begin{table}
      \caption{Basic parameters of M104}
         \label{tab:M104param}
         \begin{tabular}{lll}
           \hline
           Morphology & SA(s)a & \cite{RC3} \\
           R.A.(J2000) & 12:39:59.4& \\
           DEC (J2000) & $-$11:37:23 &          \\
           v$_\odot$ & 1095\,\kms & \cite{Tully_2016} \\
           $D_{25}$ & $8\farcm 7= 24.2\,\mathrm{kpc}$ & \cite{RC3} \\
           Distance (TRGB) & $9.55\pm0.34\,\mathrm{Mpc}$ & \cite{McQuinn_2016}*\\
           $(m-M)$ & $29.90\pm0.08$ & \cite{McQuinn_2016}\\
           $M_{B_T,0}$ & $-21.51$~mag & \cite{RC3}\\
           $v_{\rm rot}^{\rm max}$ & $345$\kms & \cite{Schweizer_1978} \\ 
           $M_{*}$ & $1.8 \times 10^{11}$~M$_{\odot}$  & \cite{Mu_oz_Mateos_2015} \\
           $M_{200}$ & $7.8 \times 10^{12}$~M$_{\odot}$ & \citep{Crosby_2023_b}\\
           $R_{200}$ & $420\,\mathrm{kpc}$\ & \citep{Crosby_2023_b} \\
           \hline
         \end{tabular}\\
         (*): Distance measurements from various techniques are given in Table 2 of \cite{McQuinn_2016}.
\end{table}

\section{Observations}

IFUM consists of three distinct fiber-optic IFUs, each feeding ‘MSpec’, a double spectrograph also used by M2FS \citep{Mateo2012}. The three IFUs are HR (High Resolution), STD (Standard Resolution) and LSB (Low surface brightness). Each of the IFUs operates by placing high-precision lenslet arrays at the focal plane of the Magellan/Clay telescope. Each lenslet is comprised of laser-etched bi-convex hexagonal surfaces in a fused-silica substrate which projects images of the telescope primary mirror (the system entrance pupil) onto the ends of precisely-aligned optical fibers (see \cite{Mateo2022} for further general details). The fibers for each IFU can be placed at the focal surfaces of the double MSpec spectrograph which was originally constructed for use with the multi-fiber, wide-field fiber spectroscopy system M2FS \citep{Mateo2012}.  The observations obtained for this study exclusively used the LSB IFU which we describe in further detail here.

\begin{figure}
	\centering
	\includegraphics[draft=false,width=8cm]{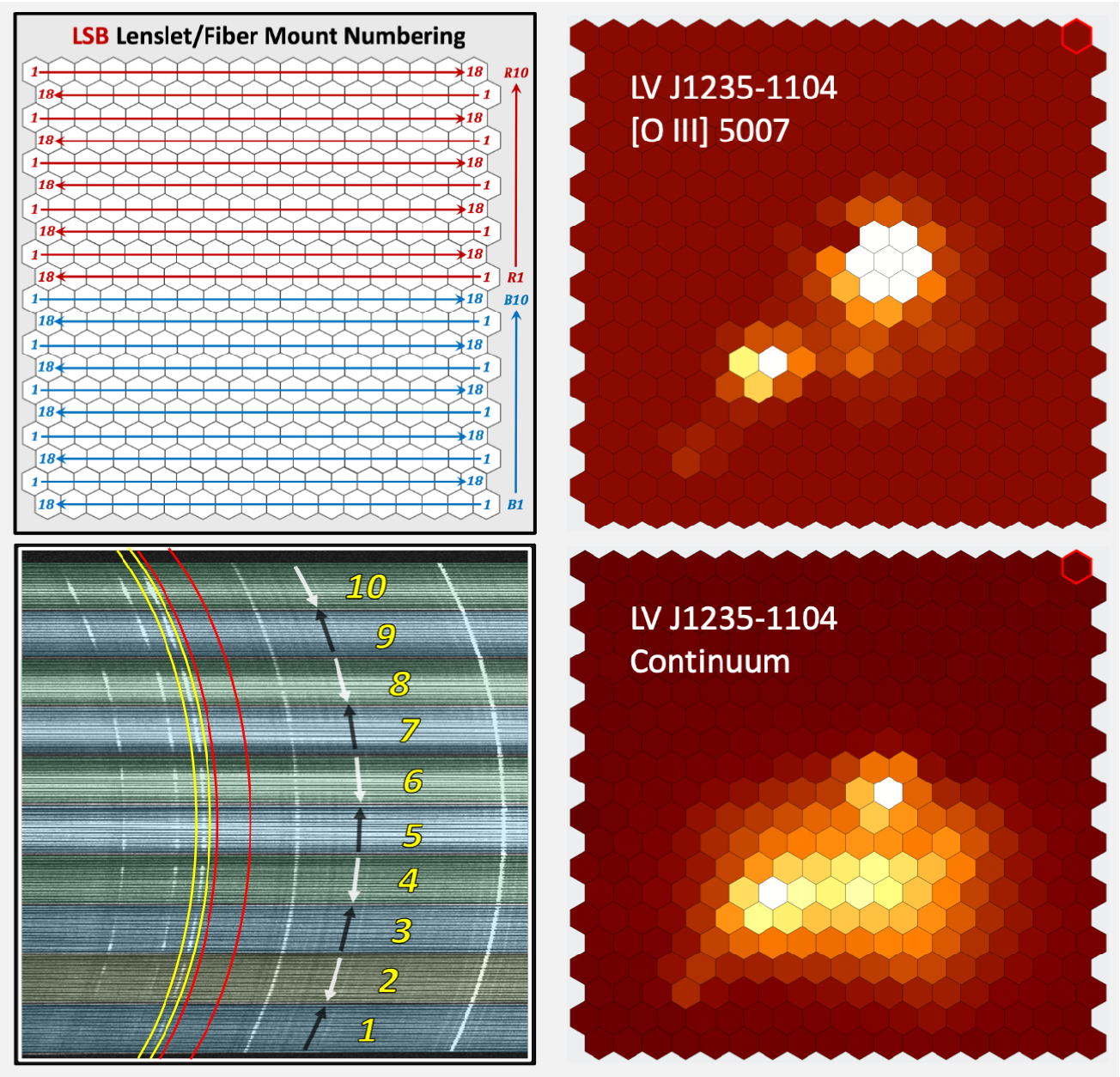}
	\caption{\textbf{(Top left)}: A schematic representation of the IFUM/LSB lenslet array as projected onto the sky with North to the top and East to the right when the Magellan/Clay Nasmyth-East rotator is at the home (zero) position. The fibers associated with the bottom half of the array feed the 'B' spectrograph while the upper half fibers feed the 'R' spectrograph (the two spectrographs are identical; the 'B' and 'R' notation is a relic from an earlier version of the instrument). The LSB fibers are mounted in groups of 18 fibers, 10 per spectrograph. The fiber mount notation and the fiber numbering is illustrated in this representation of the lenslet array. Note the zig-zag pattern of the fiber numbers.
    \textbf{(Bottom Left)}: This shows a portion (about half) of the B-side CCD observations of LV J1235-1104 (discussed further in Section \ref{sec:LVJ}). The yellow numerals refer to the 10 fiber mounts corresponding to the lower ten rows in the top left figure. Each set of 18 fibers are noted in alternating blue/green tints. The arrows show the direction of constant direction, in this case from East to West when the instrument rotator is a the home position.  Careful inspection of the galaxy emission lines (H$\beta$ and [O III] 4959/5007 at left in this image shows how features alternate in orientation in successive fiber mounts. 
    \textbf{(Upper right)}: An image of LV J1235-1104 based on the spectral region including only the [O III] emission line at 5007 \AA (between the curved yellow lines in the lower left image). 
    \textbf{(Lower right)}: An image of LV J1235-1104, this time sampling only the continuum radiation between the red arcs in the lower left image.}
	\label{fig:IFUM_format}
\end{figure}

The LSB IFU array consists of 360 individual lenslets in an $18 \times 20$ array (see Top Left panel in Fig. \ref{fig:IFUM_format}). Each spaxel of the lenslet array has an effective diameter of $1.90\arcsec$ (this equals the diameter of a circle with the same area as the hexagonal elements of the LSB IFU). The LSB IFU spans a field of $32.7\arcsec \times 31.4\arcsec$. The LSB fibers have core diameters of $260~\mu m$, and are about 3.2 m in length. Both (identical) spectrographs of MSpec are of quasi-Littrow design and hence project these fibers a $1:1$ magnification onto a pair of $4096 \times 4120$ E2V CCDs with $15~\mu m$ pixels. The two CCDs each image 180 of the fibers from half of the spaxel array. The rows of individual spaxels defined by the lenslet array are transfered, in a zig-zag pattern, to the spectrograph CCD detectors as described in Fig. \ref{fig:IFUM_format}. During the observations reported in this paper, two fibers were dead on the R-side of the LSB IFU (see Fig. \ref{fig:IFUM_format}); these fibers have since been replaced.

Data for this paper were obtained over two observing runs in March 2023 and May 2024. The Mar-2023 observations employed an $80\mu m$ wide slit just behind the output ends of the fibers at the focal surfaces of the two identical spectrographs in MSpec. These data were obtained using $1\times 2$ pixel binning (spectral × spatial) of the 4096 x 4112 CCD detectors. About 63\% of the light incident on the lenslet arrays was blocked by the 80-micron slits in this configuration. For the May-2024 run, the instrument set up was similar apart from using the 300-micron (full-open) slit which leads to no significant slit losses. The same binning ($1\times 2$) was employed as in 2023. All observations from both runs used the so-called 'blue' 600 line/mm reflection gratings in MSpec which, with the $80~\mu m$ and $300~\mu m$ slits, delivered an effective spectral resolution of $R\sim 3600$ and $R\sim 1000$ respectively over a wavelength range of 4800-6700~\AA. Table \ref{tab:IFUMtargets} indicates which slit configuration was used for each target (or just have two parts, one for LSB80 and one for LSB300). This setting comfortably spans from H$\beta$ to H$\alpha$ at the mean redshift of M104. By combining data from multiple spaxels we were able to obtain reliable absorption-line velocities of galaxies with mean surface brightnesses as faint as $24.5~{\rm mag/arcsec}^2$ in both configurations. Relevant details of the IFUM configuration used in this paper are summarized in Table \ref{tab:IFUMparam}.

Our primary science goal with these observations was to measure a reliable recessional velocity for each galaxy, coadding the spectra from each spaxel together as necessary. We eliminate the sky background from our spectra by exposing the IFU to empty sky, or a Background Target (BT) which we subtract from our Science Target (ST) frames. The total exposure time for each target was based on the galaxy’s surface brightness and was selected to produce at least a measured bulk velocity for each galaxy. Observing details and basic structural parameters of each galaxy as reported in \cite{Crosby_2023_b} are provided in Table \ref{tab:IFUMtargets}.

\begin{table}
      \caption{IFUM Instrumental Parameters}
         \label{tab:IFUMparam}
         \begin{tabular}{lllclcl}
           \hline
           & \ \ \ \ \ \ \ Parameter & & Units & & Value & \\
           \hline
           & LSB Array Size & \ \ \ \ \ \ \ & arcsec &\ \ \ \ \ \ \ & $32.7 \times 31.4$ & \\
           & LSB Array Size & & spaxels & & $20 \times 18$ & \\
           & Effective Area & & arcsec$^2$ & & 1027 & \\
           & Spaxel Diameter* & & arcsec & & 1.90 & \\
           & Slit Width & & $\mu m$ & & (80)/(300) & \\
           & Wavelength Range & & \AA & & 4800-6700 & \\
           & Effective Resolution & & $\lambda/\Delta\lambda$ & & (3600)/(1000) & \\
           \hline
         \end{tabular}\\
         (*): The spaxel effective diameter is defined as $2\sqrt{A/\pi}$ where $A$ is the nominal area of a single hexagonal spaxel of the LSB IFU. Values with format (x)/(y) indicate changes between the 2023 and 2024 observing runs respectively.
\end{table}

\begin{table*}
     \begin{tabular}
     {ccccccccccc}
      \hline           
       Name & ST time & BT time & Slit & Observing & R.A. & DEC & $\mu{_0,g}$ & $\langle\mu{_e,g}\rangle$ &  Morph& $r_e$\\
        & (s) & (s) & ($\mu m$) & Pattern & (J2000) & (J2000) & (g\,mag\,arcsec$^{-2}$) & (g\,mag\,arcsec$^{-2}$) & Type & kpc\\
       (1) & (2) & (3) & (4) & (5) & (6) & (7) & (8) & (9) & (10) & (11)\\
      \hline 
      \multicolumn{11}{|c|}{\textbf{2023 targets}} \\
        \hline 
        
        UGCA287 & 2700 & 2400 & $80$ & B-S-S-S-B & 12:33:55 & -10:40:48 & $22.2$ & $23.8$ & dIrr & 1.42\\
        LV J1235-1104 & 7200 & 4800 & $80$ & S-B-S-B-S & 12:35:39 & -11:04:01 & $18.6$ & $21.2$ & BCD & 0.28\\
        KKSG31 & 6300 & 3900 & $80$ & B-S-S-S-B & 12:38:33 & -10:29:24 & $21.3$ & $25.4$ & N-dE & 1.12\\
        dw1239-1143 & 16500 & 9600 & $80$ & S-B-S-S-B-S-S-B-S-S-B & 12:39:15 & -11:43:08 & $22.2$ & $24.4$ & N-dE & 0.73\\ 
        dw1239-1240 & 4500 & 2400 & $80$ & B-S-S-S-B & 12:39:30 & -12:40:30 & $22.2$ & $24.7$ & N-dE & 0.82\\ 
        dw1240-1118 & 13200 & 8700 & $80$ & B-S-S-B-S-S-B-S-S-B & 12:40:09 & -11:18:50 & $21.3$ & $23.8$ & N-dE & 0.72\\ 
        dw1244-1127 & 4800 & 1200 & $80$ & S-B-S & 12:44:38 & -11:27:11 & $20.6$ & $24.1$ & dIrr & 1.96\\
        \hline 
        \multicolumn{11}{|c|}{\textbf{2024 targets}} \\
        \hline 
        dw1233-0928 & 2700 & 1800 & $300$ & B-S-S-S-B & 12:33:51 & -9:28:08 & $21.1$ & $23.5$ & dIrr & 0.87\\
        PGC42730 & 2700 & 1800 & $300$ & B-S-S-S-B & 12:42:49 & -12:23:27 & $19.7$ & $23.4$ & N-dE & 1.60\\
        dw1245-1333 & 4500 & 1800 & $300$ & B-S-S-S-B & 12:45:55 & -13:32:31 & $22.9$ & $24.6$ & dE & 0.45\\

      \hline
     \end{tabular}
     \caption{Basic structural and observing parameters of the ten galaxies with IFUM spectroscopy reported in this paper. The table columns contain the following information: (1) Galaxy name, (2) total Science Target (ST) time, (3) total Background Target (BT) time, (4) slit configuration, (5) observing pattern defined with the sequence of Science Target (S) and Background Target (B) frames, (6) right ascension (J2000), (7) declination (J2000), (8) central surface brightness, (9) mean {\it g}-band surface brightness within the effective radius, (10) morphological type, and (11) half-light radius from the underlying LSB component, ignoring nuclei and star forming regions, assuming a distance to M104, 9.55$\,\mathrm{Mpc}$ \citep{McQuinn_2016}, except for dw1244-1127 which we assume the distance to NGC4680, 30.1$\,\mathrm{Mpc}$ \citep{Tully2015}. All structural parameters are from \citet{Crosby_2023_b}.}
     \label{tab:IFUMtargets}
\end{table*}

\begin{table*}
    \begin{tabular}{ccccccc}
          \hline           
          Name & R.A.& DEC& $\mu{_0,g}$ & $\langle\mu{_e,g}\rangle$ & $v_\odot$ & Reference \\
            & (J2000) & (J2000)& (g\,mag\,arcsec$^{-2}$) & (g\,mag\,arcsec$^{-2}$) & (\kms) & \\
                 (1) & (2) & (3) & (4) & (5) & (6) & (7) \\
        \hline 
            UGCA287 & 12:33:55 & -10:40:48 & $22.2$ & $23.8$ & $1052\pm9$ & \cite{RC3}\\
            LV J1235-1104 & 12:35:39 & -11:04:01 & $18.6$ & $21.2$ & $1124\pm45$ & \cite{Jones_2009}\\
            \noindent PGC042120 & 12:37:14 & -10:29:46 & $24.3$ & $25.2$ & $756\pm2$ & \cite{Huchtmeier_2009}\\
            \noindent SUCD1 & 12:40:31 & -11:40:05 & $19.4$ & $20.3$ & $1293\pm10$ & \cite{Hau_2009}\\
            \noindent PGC42730 & 12:42:49 & -12:23:24 & $19.7$ & $23.5$ & $1025\pm45$ & \cite{Jones_2009}\\
          \hline
    \end{tabular}
    \caption{Basic parameters of the five existing confirmed satellite galaxies of M104. The table columns contain the following information: (1) galaxy name, (2) Right Ascension (J2000), (3) Declination (J2000), (4) central {\it g}-band surface brightness, (5) mean \textit{g}-band surface brightness within the effective radius, (6) reference for recession velocity.}
    \label{tab:conf_satellites}
\end{table*}

\section{Results}

\subsection{Spectral Fitting}
\begin{figure}
    \begin{subfigure}{0.5\textwidth}
        \centering
        \includegraphics[height=10cm]{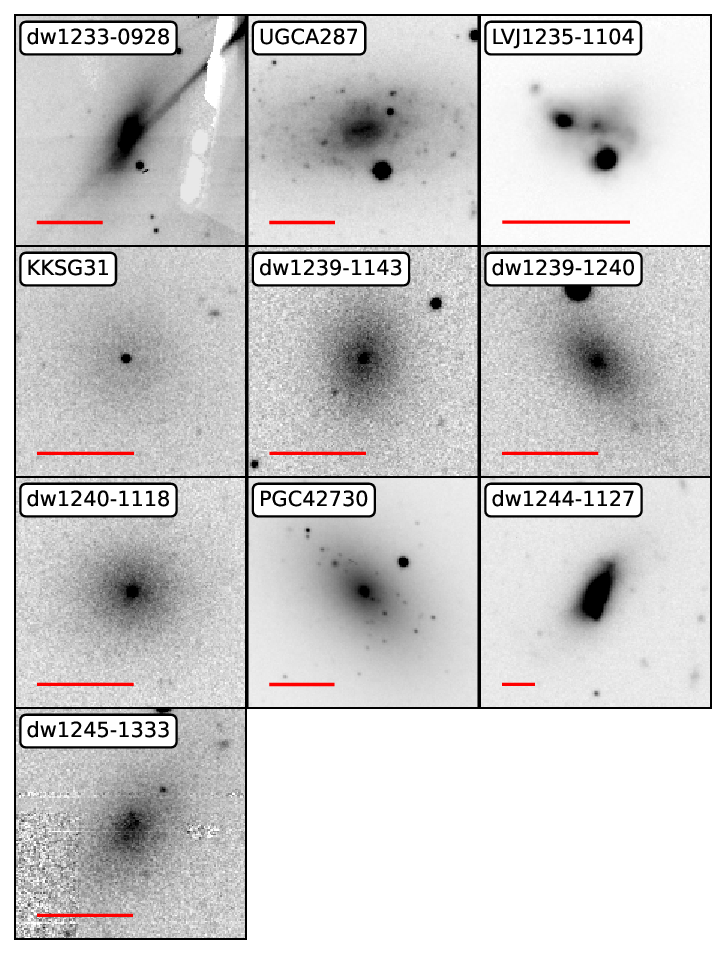}
    \end{subfigure}
    \begin{subfigure}{0.5\textwidth}
        \centering
    	\includegraphics[height=10cm]{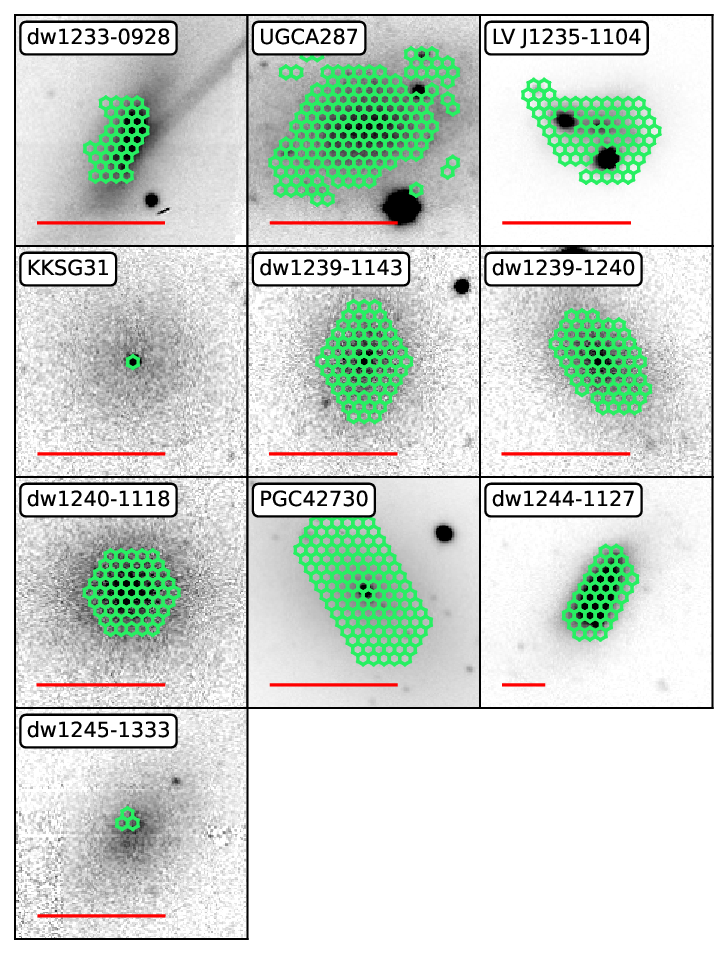}
    \end{subfigure}
    \caption{\textbf{Upper frame:} DECaLS g-band images \citep{Zou_2017,Zou_2018} of our IFUM spectroscopy targets, north is up and east to the left. The red bar in each image represents a length of $1\,\mathrm{kpc}$ at the distance of M104.
    \textbf{Lower frame:} As the Upper frame, but overlaid with hexagons indicating the spaxels considered to a part of the galaxy for the purposes of co-adding spectra to measure a bulk velocity for each galaxy. Some spaxels from galaxies are excluded when the spaxels are dead, the SNR of the spectra is too low, or it is otherwise affected by noise or foreground objects.}
    \label{fig:M104_spaxel_coadds}
\end{figure}

\begin{figure*}
	\centering
	\includegraphics[draft=false,width=18cm]{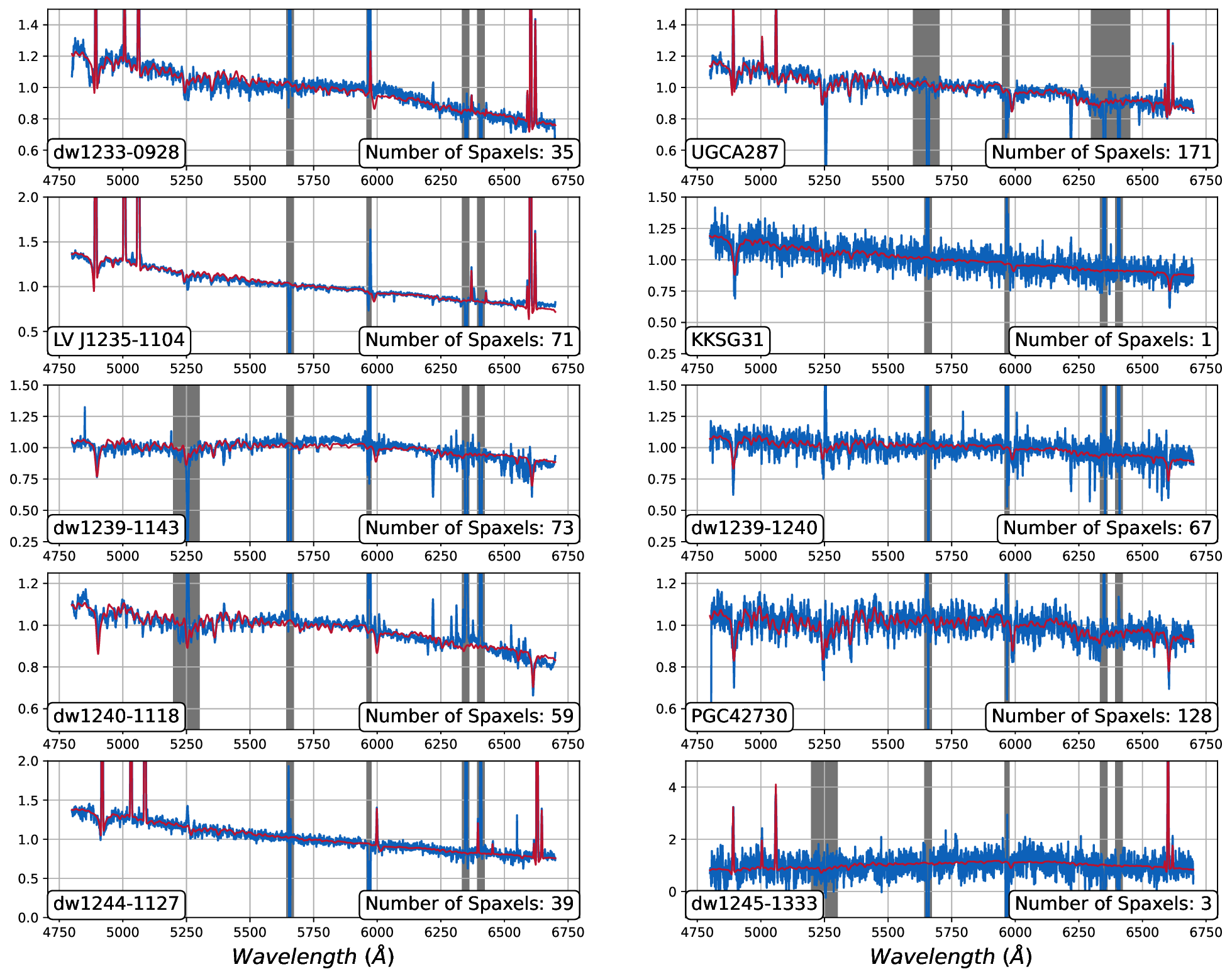}
	\caption{The co-added and continuum normalised IFUM spectrum for each target galaxy (blue line). The red line indicates the best pPXF fit from the stellar and gas templates. Greyed out windows are masked wavelength regions excluded from the fit due to the presence of saturated sky emission lines. The y-axis is an arbitrary normalised unit a function of erg\,s$^{-1}$\,cm$^{-2}\,$\AA$^{-1}$. Each panel notes how many spaxels were co-added to produce the displayed spectrum.}
	\label{fig:spectrums}
\end{figure*}

To produce our final spectra to measure systemic velocities, metallicities, weighted ages, and mass to light ratios, we co-added all spaxels of each galaxy, excluding spaxels that do not increase the signal to noise ratio when co-added, or which are clear foreground or background sources as determined typically from their radial velocities. In the unique case of LV J1235-1104, the spaxels containing the star formation regions were also excluded as unique spectral features (discussed further in Section \ref{sec:LVJ}) in these regions introduced significant fitting errors. The spaxels chosen to be co-added are shown for each galaxy in Figure \ref{fig:M104_spaxel_coadds}; the resulting IFUM spectra are shown in Figure \ref{fig:spectrums}. 

We employed the Python implementation of the Penalised PiXel Fitting (\textsc{pPXF}) algorithm \citep{Cappellari_2004,Cappellari_2016,Cappellari_2023} which employs a penalised maximum-likelihood approach to fit stellar and gas spectral templates to the IFUM data to simultaneously extract recessional velocities, stellar ages, stellar metallicities, stellar mass-to-light ratios and emission line flux ratios. \textsc{pPXF} allows one to model a galaxy spectrum with superimposed SPS models with varying weights representing the spread in age and metallicity of the underlying stellar populations, and therefore the stellar mass-to-light ratio for the given galaxy. 

For our analysis, we used the E-MILES Stellar Population Synthesis (SPS) templates\footnote{http://miles.iac.es/} \citep{Vazdekis_2016} with the following choice of parameters:
\begin{enumerate}
    \item The observation sample from the Next Generation Spectral Library (NGSL), a library of spectra for 600 stars extending from $1660-10200$\AA$\ $acquired using HST/STIS\footnote{https://archive.stsci.edu/prepds/stisngsl/index.html};
    \item Theoretical scaled solar isochrones as described in \cite{Pietrinferni2004};
    \item A bimodial $\Gamma_b=1.3$ initial mass function (IMF) resembling the IMF proposed by \cite{Kroupa2001};
    \item The assumption that [M/H]=[Fe/H], implying that there is no enhancement of $\alpha$-elements in the stellar spectra. This assumption is reasonable for dwarf elliptical galaxies similar to those observed in our program. \citep{Geha_2003,Sen2017,Sen2022}.
\end{enumerate}
Integrating the NGSL over \citep{Pietrinferni2004} isochrones and the adopted IMF generates a series of SPS models whose free parameters are the age and metallicity of the stellar population. 

For all our results, the templates are weighted by the mass of the constituent stellar population, which increases the sensitivity to older stellar populations that are less luminous than younger stellar populations but contain a greater proportion of the stellar mass of a galaxy. The alternative, using light-weighted templates, is strongly favoured towards younger stellar populations that when used to model dwarf starburst galaxies whose luminosity is dominated by young bright stars, may provide misleading mass-to-light ratios when a more massive but less luminous older stellar population is present. This is frequently the case in dwarf starburst galaxies \citep{Zhao_2011} and thus why it is essential to use mass-weighted templates for star-forming galaxies. 

We use the bootstrapping method as described in \citep{Cappellari_2023} to better estimate uncertainties in the recessional velocity, metallicity and stellar ages. This method involves perturbing the residuals of an initial singular fit to the spectra and re-fitting many times using the 'Wild' bootstrapping method \citep{Davidson2008}. We use 200 Monte-Carlo realisations for the results of each galaxy reported here. Additionally, we also use a regularisation parameter of 3 for each galaxy to further smooth the results.

To calculate the stellar mass, we use the \textit{i} and \textit{r} band magnitudes as measured from DECaLS {\it i} and {\it r}-band images \citep{Zou_2017,Zou_2018}, assuming the distance to M104 (or NGC4680 in the case of dw1244-1127) to calculate the luminosities and thus the stellar mass from the aforementioned mass-to-light ratio. We use the {\it i} and {\it r}-bands as it is less influenced by internal extinction and strong emission lines in star-forming galaxies than bluer pass bands.

An additional component that is sometimes required to model a galaxy's spectrum is the gas component, which, if present, can produce observable emission features if it is excited by energetic sources, eg. star-forming regions. For this study, we have constructed a emission spectral template that contains the following strong emission lines found in the observed spectral range:
\begin{enumerate}
    \item H$\alpha$ $\lambda6563$ and H$\beta$ $\lambda4861$ Balmer series lines
    \item $\left[\textsc{OIII}\right]$ $\lambda4959$ and $\lambda5007$ doublet
    \item HeI $\lambda5876$
    \item $\left[\textsc{OI}\right]$ $\lambda6300$ and $\lambda6369$ doublet
    \item $\left[\textsc{NII}\right]$ $\lambda6548$ and $\lambda6584$ doublet
\end{enumerate}
The relative ratios of these lines are allowed to vary as often seen in spectra of star forming galaxies. We use the measured ratios of the H$\beta$ and H$\alpha$ emission features to estimate the gas attenuation, from the Balmer decrement and case B recombination \citep{Storey_1995}. Using equations 23a, b and c from \cite{Cappellari_2023} for a generic extinction curve, we can thereby estimate the attenuation at any wavelength in our spectra.

To calibrate our spectra to physical units, we first convolve the DECam r-band filter with the optimal pPXF fit for the dwarf ellipticals dw1239-1143, dw1239-1240 and dw1240-1118 to simulate a DECam r-band magnitude. Then, we used the absolute magnitude as calculated from DECaLS r-band images \citep{Zou_2017,Zou_2018} as described above. The Star Formation Rate (SFR) can be estimated from the calibrated H$\alpha$ intensity \cite{Calzetti_2007}:
\begin{equation}
    \mathrm{SFR}(M_{\odot}\,\mathrm{yr}^{-1}) = 5.3\times10^{-42}\,L(\mathrm{H}\alpha)_{\mathrm{corr}}\,(\mathrm{erg}\,\mathrm{s}^{-1})
\end{equation}
Where $L(\mathrm{H}\alpha)_{\mathrm{corr}}$ is correct for extinction. The SFR for each galaxy is provided in Table \ref{tab:IFUMResults}, as calculated from the co-added flux from all spaxels containing a $\mathrm{H}\alpha$ signal.

For each galaxy spectra, we calculate the Signal to Noise Ratio (SNR) using the following equation:
\begin{equation}
    \mathrm{SNR} = \langle \mathrm{ST}(\text{\AA})\rangle\,/\,\sqrt{\langle\mathrm{BT}(\text{\AA})\rangle\,*\,\chi}
\end{equation}
Where ST and BT represent the Science Target and the Background Target frames respectively. $\chi$ comes from the pPXF fitting parameters and represents the quality of the fit. To avoid the influence of emission and absorption features, we select a 200\AA$\ $region from about $5950-6150$\AA$\ $ that is largely free of spectral features to make this comparison, thus effectively measuring the SNR of the continuum signal. The SNR for each galaxy is provided in Table \ref{tab:IFUMResults}.

\begin{figure*}
	\centering
	\includegraphics[draft=false,width=18cm]{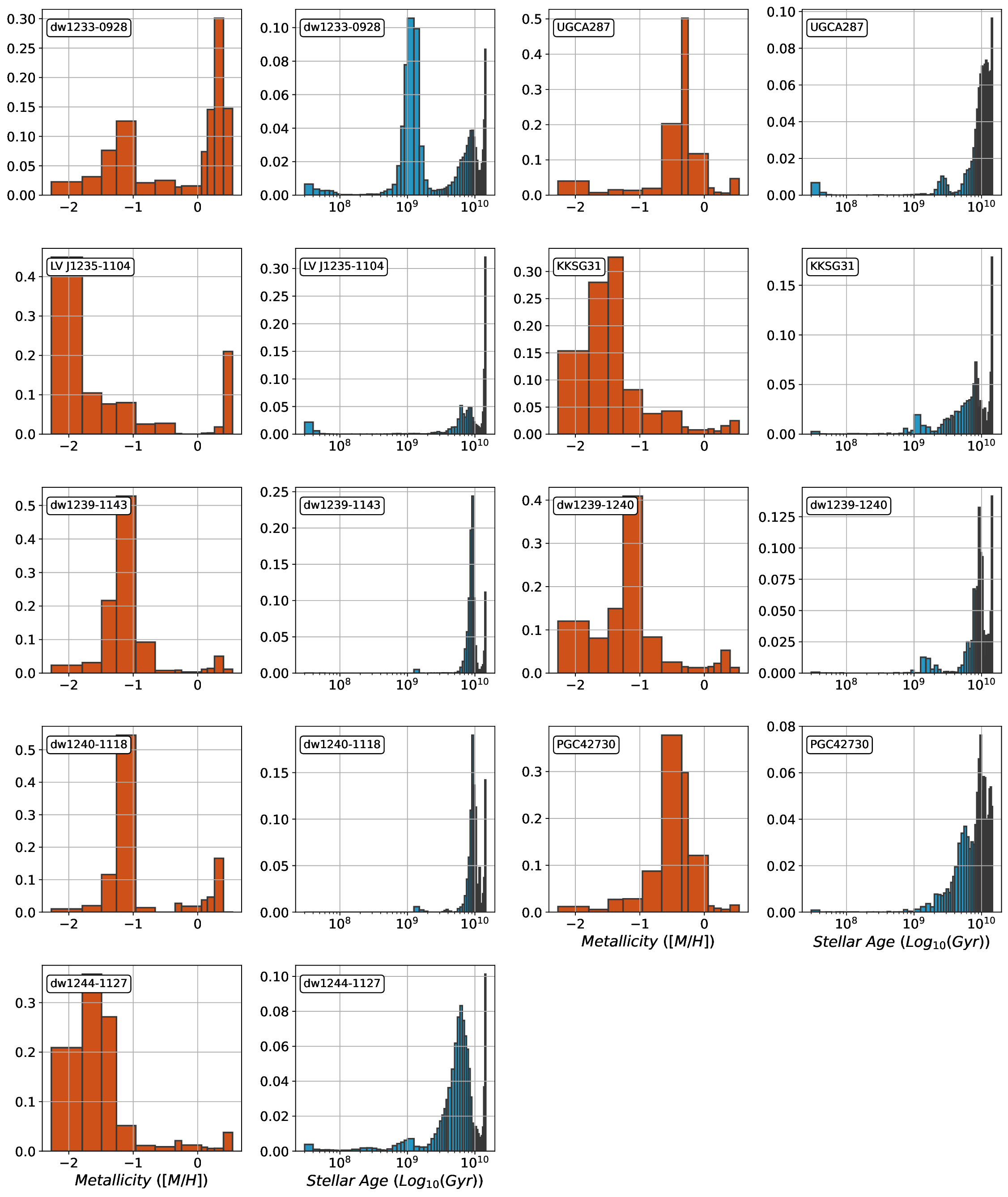}
	\caption{The normalised weights for the metallicity and stellar age of the E-MILES SPS templates fit to the data IFUM galaxy spectra by pPXF.}
	\label{fig:age_metal}
\end{figure*}

\begin{table*}
         \begin{tabular}{ccccccccc}
          \hline           
            Name & $\langle{\rm age}\rangle$ & $\langle{[M/H]}\rangle$ & H$_\alpha\,$SFR & $v_{\odot}$ & $\Delta{v}_{\odot,M104}$ & $M_{*}$ & \# spaxels & SNR\\
             & (Gyr) & (dex) & ($M_\odot$\,yr$^{-1}$) & (\kms) & (\kms) & ($10^{7}\,\times\,M_\odot$) & & \\
                       (1) & (2) & (3) & (4) & (5) & (6) & (7) & (8) & (9)\\
          \hline 
          \multicolumn{9}{|c|}{\textbf{2023 targets}} \\
          \hline 

            UGCA287 & $9.3^{+1.3}_{-4.2}$ & $-0.5^{+0.2}_{-0.2}$ & 0.0013 & $1001\pm1$ & $-94\pm5$ & $53^{+6}_{-7}$ & 171 & $38.7$\\
            LV J1235-1104 & $8.5^{+2.7}_{-2.4}$ &  $-1.7^{+0.6}_{-0.4}$ & 0.042 & $1074\pm1$ & $-17\pm5$ & $8.4^{+2.7}_{-1.8}$ & 71 & $61.3$\\ 
            KKSG31 & $6.5^{+3.1}_{-1.9}$ & $-1.4^{+0.3}_{-0.3}$ & 0.0 & $1309\pm22$ & $214\pm23$ & $2.8^{+0.7}_{-0.6}$ & 1 & $2.2^{(*)}$\\
            dw1239-1143 & $9.0^{+0.8}_{-0.7}$ & $-1.2^{+0.1}_{-0.1}$ & 0.0 & $1367\pm12$ & $272\pm13$ & $4.9^{+0.7}_{-0.4}$ & 73 & $18.3^{(*)}$\\ 
            dw1239-1240 & $8.7^{+1.6}_{-1.3}$ & $-1.2^{+0.3}_{-0.3}$ & 0.0 & $1040\pm13$ & $-55\pm14$ & $7.9^{+0.5}_{-0.5}$ & 67 & $7.6^{(*)}$\\
            dw1240-1118 & $9.0^{+0.9}_{-1.0}$ & $-0.9^{+0.1}_{-0.1}$ & 0.0 & $1593\pm9$ & $498\pm10$ & $8.0^{+0.9}_{-0.3}$ & 59 & 31.8\\ 
            dw1244-1127 & $5.4^{+2.7}_{-1.1}$ & $-1.3^{+0.9}_{-0.6}$ & 0.035 & $2441\pm1$ & $1346\pm5$ & $29^{+15}_{-5}$ & 39 & $16.2^{(*)}$\\
          \hline 
          \multicolumn{9}{|c|}{\textbf{2024 targets}} \\
          \hline 
            dw1233-0928 & $2.7^{+2.3}_{-1.1}$ & $-0.3^{+0.4}_{-0.5}$ & 0.0016 & $1137\pm2$ & $42\pm6$ & $9.8^{+4.9}_{-3.4}$ & 35 & 36.2\\
            PGC42730 & $10.9^{+1.3}_{-0.8}$ & $-0.5^{+0.1}_{-0.1}$ & 0.0 & $1154\pm4$ & $59\pm7$ & $98^{+8}_{-7}$ & 128 & 50.5\\
            dw1245-1333 & - & - & $3.6\times10^{-5}$ & $981\pm1$ & $-114\pm5$ & - & 3 & $0^{(*)}$\\
          \hline
         \end{tabular}
         \caption{Kinematic and stellar population parameters of the ten satellite galaxy candidates derived from galaxy spectrum. The columns contain the following information: (1) Galaxy name, (2) mass-weighted stellar age, (3) mass-weighted stellar metallicity, (4) star formation rate calculated from H$_\alpha$ luminosity, (5) heliocentric velocity, (6) heliocentric velocity difference between the satellite and M104, (7) stellar mass, (8) number of spaxels used in building the galactic spectra shown in Fig.\ref{fig:spectrums}, (9) continuum Signal to Noise Ratio of each resulting spectra. The heliocentric velocity of M104 is $1095\pm5$\,\kms \citep{Tully_2016}. The weighted age and metallicity are calculated from the weighted sum of the histograms in Fig. \ref{fig:age_metal}. (*): The SNR of these spectra are <20, thus the mean age and metallicities provided may be biased \citet{Woo_2024}.}
         \label{tab:IFUMResults}
\end{table*}

\subsection{Satellite Membership} \label{sec:sat_membership}
Critical to testing the hypotheses central to our science goals is confirming the membership of the dwarf galaxies presented in the paper as satellites of M104. A quantitative method to achieve this involves modelling the escape speed of the M104 halo as a function of radius from its centre. The LOS velocity of any satellite that significantly exceeds the escape speed at a given projected radius can be confidently excluded as a satellite of the M104 system.

To carry out this analysis, we assume that the M104 halo is spherically symmetric and that it can be reasonably modeled with an Einasto potential, $\phi_{e}(\textsc{r})$, as formulated by \cite{RetanaMontenegro2012} (see eq. 19).

\begin{figure}
	\centering
	\includegraphics[draft=false,width=8cm]{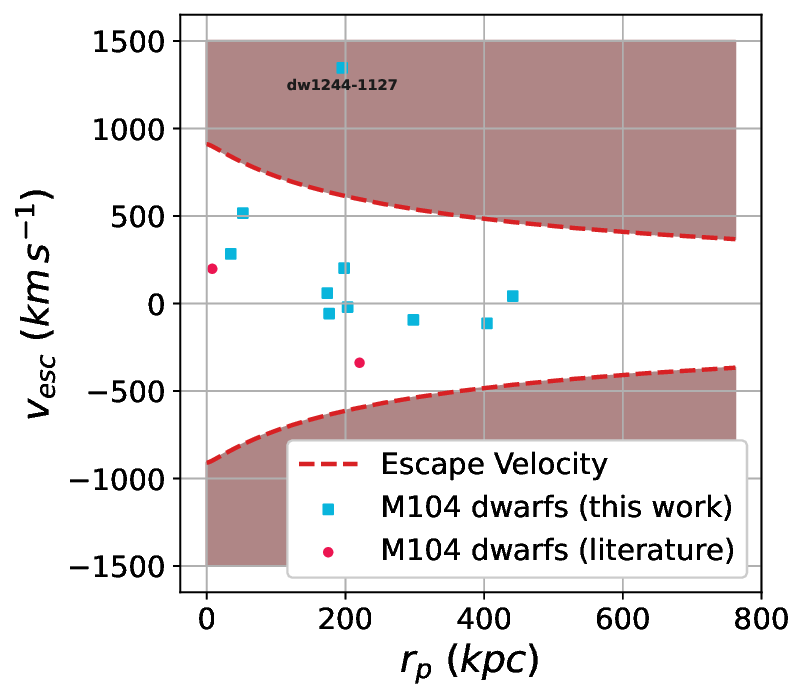}
	\caption{The escape velocity from the M104 halo ($v_{esc}$) as a function of projected radius ($r_{p}$) plotted with a dashed line. All spectroscopically measured M104 satellite velocities in this paper are plotted with squares, while velocities sources from elsewhere in the literature are represented with circles. dw1244-1127 is labelled separately in the plot and lies inside the shaded area which exceeds the escape velocity of the M104 system, and thus is not a satellite of M104.}
	\label{fig:M104_escape_vel}
\end{figure}

Assuming that the M104 halo is spherically symmetric, we model the gravitational potential of the M104 system using the Einasto potential as formulated in Equation 6b from \cite{Miller2016}:
\begin{equation}
    \phi(r) = \frac{-GM}{r}\left[ 1-\frac{\Gamma \left( 3n,{\frac{r}{r_0}}^{\left(1/n\right)}\right)}{\Gamma\left(3n\right)} +\frac{r}{r_0}\frac{\Gamma \left( 2n,{\frac{r}{r_0}}^{\left(1/n\right)}\right)}{\Gamma\left(3n\right)}\right]
\end{equation}
We have adopted an Einasto index of $n=3$ and a scale radius of $250\,\mathrm{Mpc}$.  These parameters provide a rotation curve that is a reasonably flat extrapolation of the inner rotation curve \citep{Faber1977,Lewis1985} out to the virial radius of M104 and given the mass from Section \ref{dyn_mass} below. We acknowledge the circular logic involved wherein by the mass of M104 is influenced by the choice to include certain dwarf galaxies in the method in Section \ref{dyn_mass}, but the mass of M104 calculated here is consistent with independent measurements, such as in \cite{Karachentsev2020_b} and the satellite membership of the dwarf galaxies is not strongly affected by variation in these parameters. Given the potential $\phi_{e}(\textsc{r})$, the escape velocity as a function of radius is:
\begin{equation}
    v_{esc}(r) = \sqrt{2*\phi_{e}(\textsc{r})}
\end{equation}
We plot the escape velocity cone as a function of radius in Figure \ref{fig:M104_escape_vel}, along with the the recessional velocities of M104 dwarf satellite candidates including those from this paper. We find that with the exception of dw1244-1127 (which as illustrated in Figure \ref{fig:M104_satellite_map} is most likely a satellite of the background host galaxy NGC4680), that all the dwarf candidates are realistic satellites of M104 as they lie within the escape velocity cone.

\subsection{Individual Galaxies}



\subsubsection{KKSG31}

KKSG31 is a nucleated dwarf elliptical for which we were only able to extract a reliable spectrum from the nuclear star cluster (NSC, Fig. \ref{fig:M104_spaxel_coadds}) from which we obtain a recessional velocity $v_\odot=1297\pm7$\,\kms. We measure a mean mass-weighted metallicity of $\langle[M/H]\rangle=-1.4$ and a mean weighted age of 6.5$\,$Gyr, within expectations for dwarf elliptical galaxies of this luminosity. The low S/N of 2.2 in the spectrum of KKSG31's NSC means we can't fully resolve the relevant metal absorption features (MgB ($\lambda5205$ and $\lambda5193$) and NaD ($\lambda5980$ and $\lambda5985$)). The spread in metallicities shown in Figure \ref{fig:age_metal} for KKSG31 is the result of an upper limit on the intensity on those metal lines.

\subsubsection{dw1245-1333}
dw1245-1333 is the only non-nucleated dE in our sample and hence has too low a SNR for a reliable estimate of its population parameters from its continuum spectrum. Thus we do not report stellar metallicities or ages for this galaxy. However, we did detect a weak emission signal in this galaxy localised to a small region offset from the galactic centre (as shown by the spaxel selection in Fig. \ref{fig:M104_spaxel_coadds}), whose spectra suggests it is a small star forming region. We used the emission signal from this star formation region to measure a recessional velocity of $v_\odot=981\pm1$\,\kms for this galaxy. The presence of this star forming region might better classify this galaxy as a transitional dwarf galaxy (dT) or a dwarf irregular (dIrr).

\subsubsection{dw1240-1118, dw1239-1143, dw1239-1240 and PGC42730}
dw1240-1118, dw1239-1143, dw1239-1240 and PGC42730 are all nucleated dEs, possessing Nuclear Star Clusters (NSC). Interestingly, dw1240-1118 and dw1239-1143 both have higher recession velocities, of $1617\,$\kms and $1378\,$\kms respectively, which is redshifted relative to M104 by $516\,$\kms and $283\,$\kms. However, the escape velocity cone in Figure \ref{fig:M104_escape_vel} suggests these galaxies are still bound to M104. We conclude that all 3 galaxies are satellites of M104. 

\begin{figure}
	\centering
	\includegraphics[draft=false,width=8cm]{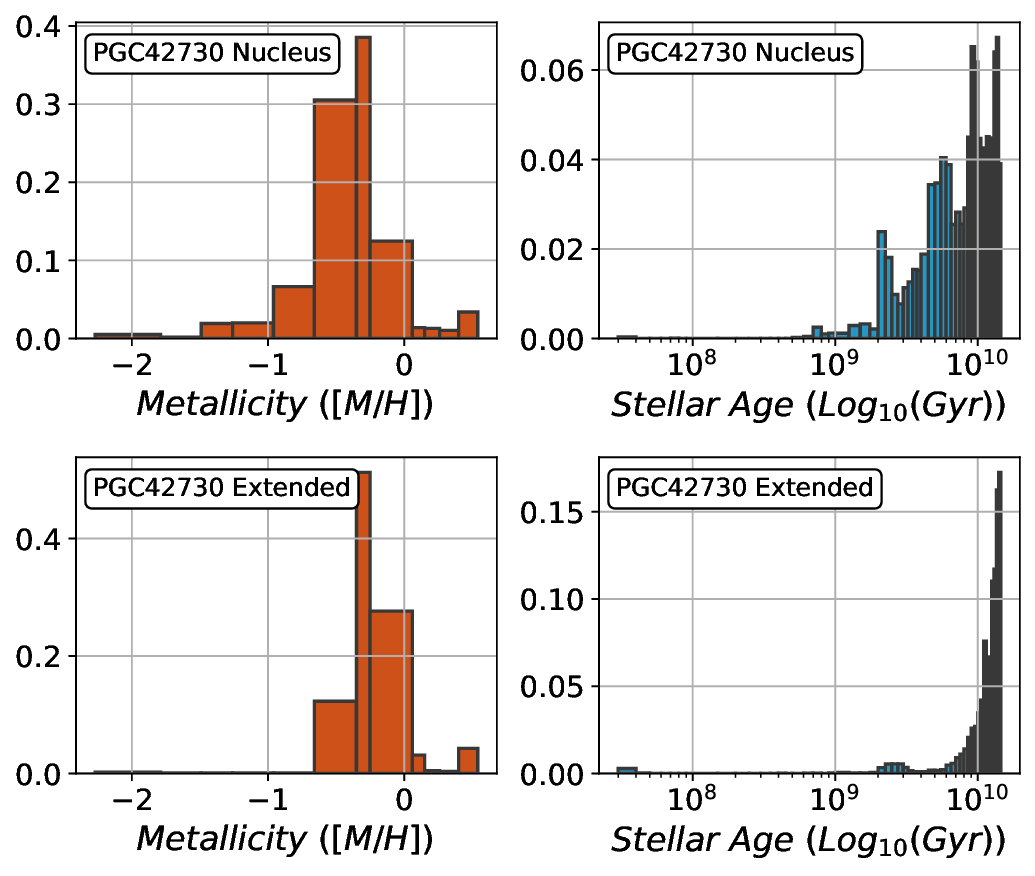}
	\caption{The stellar metallicity and age parameters for the stellar population of the nucleus (top row) and the rest of the galaxy, or the extended component (bottom row) for the nucleated dwarf elliptical PGC42730. The ongoing star formation up to $\sim1\,\textsc{Gyr}$ ago in the nucleus suggests an in-situ formation scenario for this nuclear star cluster.}
	\label{fig:PGC42730_nucleus}
\end{figure}

The SNR for PGC42730's nucleus and surrounding galaxy spectra (18.9 and 46.9 respectively) are high enough that we can identify variations in the  star formation history within this galaxy. As shown in Fig. \ref{fig:PGC42730_nucleus}, while both the nucleus and the rest of the galaxy have similar metallicity, a continuous range of younger stars in the nucleus suggests a scenario where ongoing star formation built up the nucleus in-situ until around $\sim1\,$Gyr ago. The stellar mass of the nucleus is $M_{*}=1.1\pm0.2\times10^7\,M_{\odot}$, which means the fraction of stellar mass that resides in the nucleus is $0.01-0.02$. This is expected for a galaxy of PGC42730's stellar mass, and matches the properties of nuclear star clusters found in other galaxies of this mass \citep{Fahrion2022}.

\subsubsection{UGCA287 and dw1233-0928}

UGCA287 and dw1233-0928 are both typical dwarf irregular galaxies with recessional velocities $v_\odot=1000\pm1$\,\kms and $v_\odot=1137\pm2$\,\kms respectively. These galaxies are both large, with effective radii $r_{e}\sim0.5$ and $r_{e}\sim0.33$ arcmininutes respectively ($1400\,\mathrm{pc}$ and $900\,\mathrm{pc}$ at the distance of M104), parts of the galaxy extends beyond the field of view of IFUM. dw1233-0928 consists of 2 constituent stellar populations split into a bimodal metal rich and metal poor population (as shown in Figure \ref{fig:age_metal}). These dwarf irregulars have star formation rates of 0.0013 and 0.0016 $M_\odot$\,yr$^{-1}$ respectively, but the star formation rate for UGCA287 may be underestimated as all star forming regions were not captured within the IFUM field of view.

\subsubsection{LV J1235-1104} \label{sec:LVJ}

LV J1235-1104 is a low mass blue compact dwarf (BCD) ($M_{*}=8.4\times10^7\,M_{\odot}$) and a candidate Green Pea (GP) \citep{Cardamone_2009} galaxy. The classical Green Peas as originally reported in \cite{Cardamone_2009} have stellar mass $M_{*}\sim10^{8.5}-10^{10}\,M_{\odot}$, making a LV J1235-1104 a low-mass Green Pea candidate, that some call a 'Blueberry' \citep{Yang2017,Liu2022}. BCD and GP galaxies are both rare classes of galaxies \citep{Lee2009,Karachentsev2020_a} that are effectively dwarf starburst galaxies. Both classes are described purely by the colours and shapes of the galaxies in telescope images, both being compact and blue or green in colour, indicative of strong [\ion{O}{III}] $\lambda 5007$\AA \,emission \citep{Cardamone_2009}. Indeed, this emission line within LV J1235-1104's integrated spectra has a very large equivalent width of 314$\,$\AA$\ $ which falls above the threshold of an Extreme Emission Line Galaxy (EELGs) as defined in \cite{Amorin_2015} (EW[\ion{O}{III}]$>100\,$\AA), which as a broad category include GP galaxies. EELGs are dominated by young stellar populations with high specific star formation rates (sSFR$\sim 1-100\times 10^{-9}$\,yr$^{-1}$) and low mean metallicity (log$[O/H]+12\sim8$) \citep{Yang2017}. Thanks to recent discoveries from James Webb Space Telescope research programs these galaxies have been shown to closely resemble the first galaxies of the universe \citep{Curti_2022,Harish_2022,Taylor_2022,Rhoads_2023,Trump_2023}. Curiously, while the EW of the emission features in LV J1235-1104 are indicative of extreme star forming conditions akin to these high redshift dwarf starburst galaxies, its sSFR ($sSFR\sim5*10^{-10}\,yr^{-1}$) is lower than the typical definition of an EELG. Modern ($z\sim 0$) starburst dwarf galaxies are distinct from high-z galaxies in that they often contain a population of older stars \citep{Amorn2007,Amorn2012} which contains $>90\%$ of the stellar mass of the galaxy. Earlier galaxies, which have not yet had the time to form an older population of stars, might therefore have a systematically higher sSFR, even among galaxies of comparable star forming conditions and thus may account for the unexpectedly low sSFR in LV J1235-1104.

\begin{figure}
	\centering
	\includegraphics[draft=false,width=8cm]{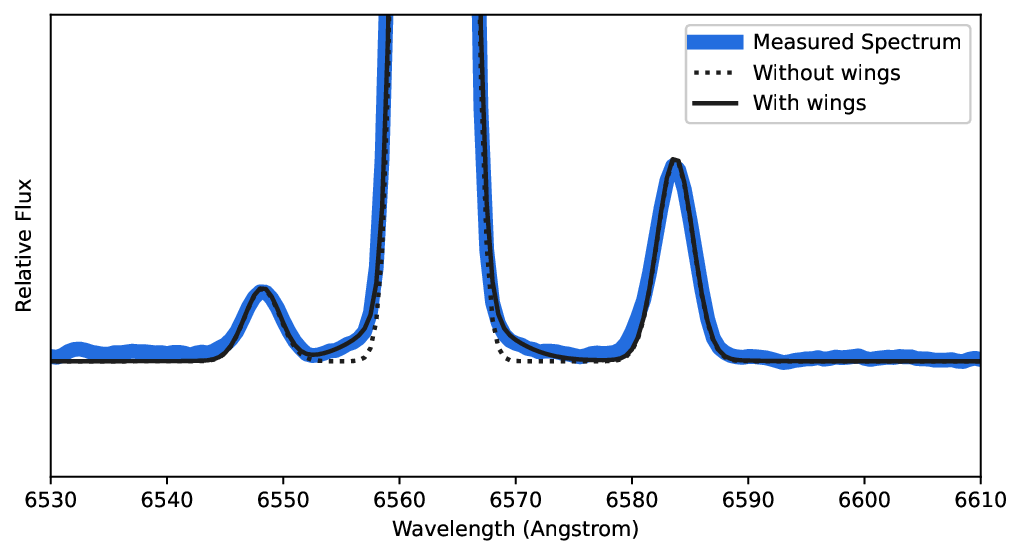}
	\caption{The line profiles of the Balmer H-alpha transition and nearby Nitrogen and Oxygen forbidden transitions from the co-added spectra of LVJ1235-1104, shown with the tricker, transparent line. The fits to all three emission features are shown without gaussian wings (dotted line) and with gaussian wings (solid line). The 'wing' component is consistent with an ionised galactic wind with velocity $\sim250\,\mathrm{km\,s^{-1}}$}
	\label{fig:BCD_line_wings}
\end{figure}

Additionally, recent IFU observations of GPs show they exhibit unique characteristics. They are intensely star forming galaxies and are frequently Lyman Continuum (LyC) and Lyman Alpha (Ly$\alpha$) emitters \citep{Izotov_2017,Izotov_2018,Komarova_2021,Malkan2021,Flury_2022} and have a broad line component of spectral emission features with $\sigma\sim100-300$\,\kms \citep{Bosch2019,Hogarth_2020,Komarova_2021}. We have confirmed the presence of similar broad emission features in this galaxy as shown Fig. \ref{fig:BCD_line_wings}. These features originate from excited gas accelerated to high velocities. The source of this acceleration must be related to star formation, likely the radiation pressure from super-star clusters, which is expected to dominate over thermal pressure \citep{Krumholz_2009}. Secondly, \cite{Komarova_2021} ruled out a scenario where radiation pressure acting on the dust grains is responsible for the observed high velocity wind: the dust opacity must be around 2 orders of magnitude higher than even in a solar metallicity system in order to accelerate the gas to these high velocities. Among the only absorption mechanisms which has enough opacity to produce these high winds is Ly$\alpha$/LyC absorption of neutral hydrogen. \cite{Komarova_2021} concluded that provided Ly$\alpha$/LyC photons are able to escape the star forming regions, this is a viable explanation for these high velocity, star-formation driven winds. This provides evidence that LV J1235-1104 is a Ly$\alpha$/LyC emitter like higher redshift GPs, but further observations such as direct UV spectroscopy with HST/COS is required to confirm this hypothesis.

\begin{figure}
	\centering
	\includegraphics[draft=false,width=8cm]{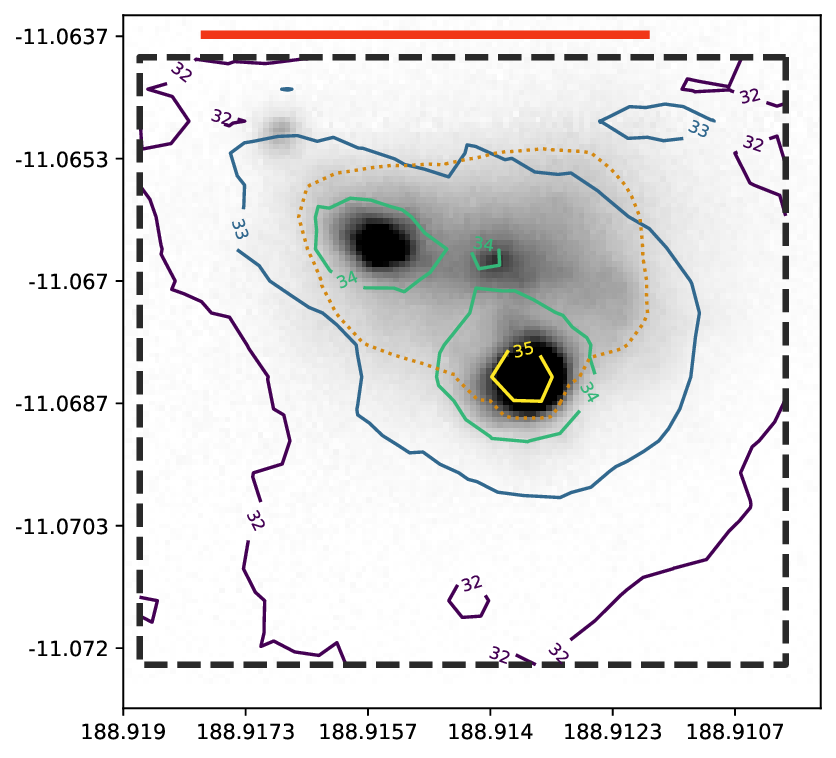}
	\caption{A contour plot of the H$\alpha$ emission intensity across the IFUM field of view, bounded by the thick dashed line. The horizontal thick line above the IFUM FOV represents a distance of $1\,\mathrm{kpc}$ at the distance of M104. The numbers superimposed on the contours represent the H$\alpha$ emission surface brightness with units of log($\mathrm{erg}\,\mathrm{s}^{-1}\,\mathrm{pc}^{-2}$), assuming the distance to the galaxy is $9.55\,\mathrm{Mpc}$. The dotted contour represents the area containing the $>99\%$ of the stars within this galaxy.}
	\label{fig:BCD_Ha_cont}
\end{figure}

We also detect the presence of a star-formation powered outflow of hot, ionised gas from an extended envelope of optical emission features in the circum-galactic medium (CGM). In Figure \ref{fig:BCD_Ha_cont} we show the H$\alpha$ emission surface brightness contour plot, which reveals the extended emission envelope that extends beyond the Stellar component of the galaxy, indicating that excited gas is being ejected from the star forming regions into the CGM. Galactic outflows with outflow velocities that exceed the escape velocity of the host halo are common in low-mass BCD galaxies \citep{Romano_2023}.

Low mass dwarf galaxies have low rotational velocities and low escape speeds, such that feedback resulting from star formation or environmental interactions such as ram pressure stripping are thought to remove their star forming gas reserves rapidly \citep{Wetzel2015,Simpson2018,GarrisonKimmel2019_b,Putman_2021}. This is supported with observational evidence, star forming dwarf galaxies are much more common in the field, rather than inside of satellite galaxy systems, suggesting they are rapidly quenched in the presence of $L_*$ host galaxies or that the mechanisms powering their intense star formation is strongly suppressed in satellite galaxy environments. Interactions between low mass dwarfs and an $L_*$ host galaxy quench starburst episodes as opposed to triggering them \citep{Brunker2022,Laufman_2022}. LV J1235-1104 is both likely experiencing strong internal star formation feedback and environmental stripping from M104 (a projected distance of $200\,\mathrm{kpc}$ to M104) which given its low mass indicates it will soon be stripped of its star forming gas. Additionally, given its very low relative velocity ($\Delta v_{\odot,M104}=-20\,\mathrm{km\,s^{-1}}$), it must either be orbitting M104 in the plane of the sky or near it's orbital apoapsis. Given that this galaxy still possesses its star forming gas, the later scenario is likely, indicating that this galaxy is on its first infall into the M104 group.

For the results presented in Table \ref{tab:IFUMResults} we deliberately exclude the spectra from these star formation regions. The significant strength of typically weak emission lines (such as [\ion{Cl}{III}] $\lambda 5517$\AA and [\ion{Cl}{III}] $\lambda 5537$\AA) and broad line features introduced significant fitting errors when attempting to fit the spectra with \textsc{pPXF}, since these features are not accounted for by \textsc{pPXF}.

\subsubsection{dw1244-1127}

dw1244-1127 is a star-forming dwarf irregular galaxy with a recessional velocity of $v_\odot=2441\pm1$\,\kms, placing it far in background behind M104, and thus isn't a member of the M104 group. As shown in Figure \ref{fig:M104_satellite_map}, this galaxy is likely a satellite of the nearby spiral galaxy NGC4680, with a velocity difference of $\Delta{v}=25\pm1$\,\kms.

\subsection{Coherent Satellite Motion} \label{satellite_plane}

\cite{Crosby_2023_b} hypothesised that LOS velocities alone may be sufficient to indicate the presence of a kinematic satellite plane, by comparing the mean value and width of the satellite LOS velocity distribution. If a stable plane consisting of coherently rotating satellites was observed at least partially face-on, then the expectation is that the LOS velocity distribution would appear abnormally cold for an environment of M104's mass, since the motions of each satellite would then be largely in the on-sky proper motions, rather than the LOS motions. Currently we lack the means to acquire on-sky proper motions to directly reproduce the kinematics and determining highly precise (uncertainties less than $\pm100\,\mathrm{kpc}$) distances to satellites is difficult at the edge of the Local Volume where M104 resides. 

\begin{figure}
	\centering
    \includegraphics[draft=false,width=8cm]{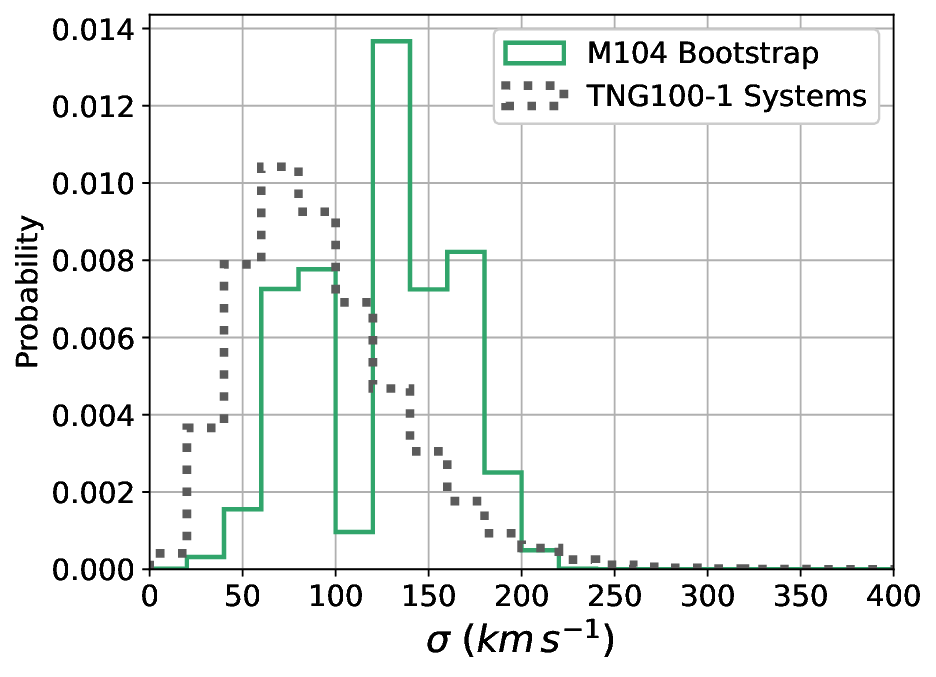}
	\caption{
    The distribution of the standard deviation of relative line-of-sight velocities for the bootstrapped sample of the M104 satellites with known recessional velocities (solid line) and the 265 similar M104 environments from the TNG100-1 simulation (dotted line). The mean velocity dispersions are $\sigma=138\pm36\,\mathrm{km\,s^{-1}}$ and $\sigma=92\pm43\,\mathrm{km\,s^{-1}}$ for the bootstrapped sample of M104 satellites and the M104 analogue systems in TNG100-1 respectively.
    }
	\label{fig:satellite_dispersions}
\end{figure}

We start by measuring the velocity dispersion of the likely M104 satellite members (section \ref{sec:sat_membership}) that have reliable radial velocities. To estimate the uncertainty in these velocities, we bootstrap the results, by randomly sub-sampling satellites with measured velocities allowing for replacement. Then, we compare these results to simulated M104 analogues from the cosmological simulation Illustris: The Next Generation 100 (Illustris TNG-100) \citep{Springel_2017,Nelson_2017,Naiman_2018,Marinacci_2018,TNG_MAIN,Nelson_2019}. A total of 265 M104-like analogues were selected this way, as defined in \cite{Crosby_2023_b}. We then selected groups using a Friends Of Friends (FOF) groups that are separated from each other by at least 2 virial radii ($R_{200}$) and are selected based on the virial mass ($M_{200}=10^{12}-10^{13}\,M_{\odot}$) and the number of sub-halos ($n_{sub}=30-100$) within the virial radius with at least an absolute g-band magnitude of $M_g=-9$. For each analogue, we select the 10 brightest satellites in the g-band (satellites as selected in \cite{Crosby_2023_b}) and perform this same bootstrapping method to produce a velocity dispersion distribution for each simulated analogue, simulating a mock set of observations in a similar manner presented here for M104. All these distributions are then added together. We compared the LOS velocity dispersion distributions from simulated M104 analogues and M104 itself in Figure \ref{fig:satellite_dispersions}. From this process, we measure a mean satellite dispersion velocity of $\sigma=92\pm43$\,\kms\,for M104-like analogues in Illustris TNG-100, and $\sigma=138\pm36$\,\kms\,from our bootstrapped M104 satellites. Both distributions nearly lie within the $1\,\sigma$ of each other and thus we are unable to conclude that there exists the presence of unexpected coherent motion in the satellites in comparison to simulations.

\begin{figure*}
	\centering
	\includegraphics[draft=false,width=16cm]{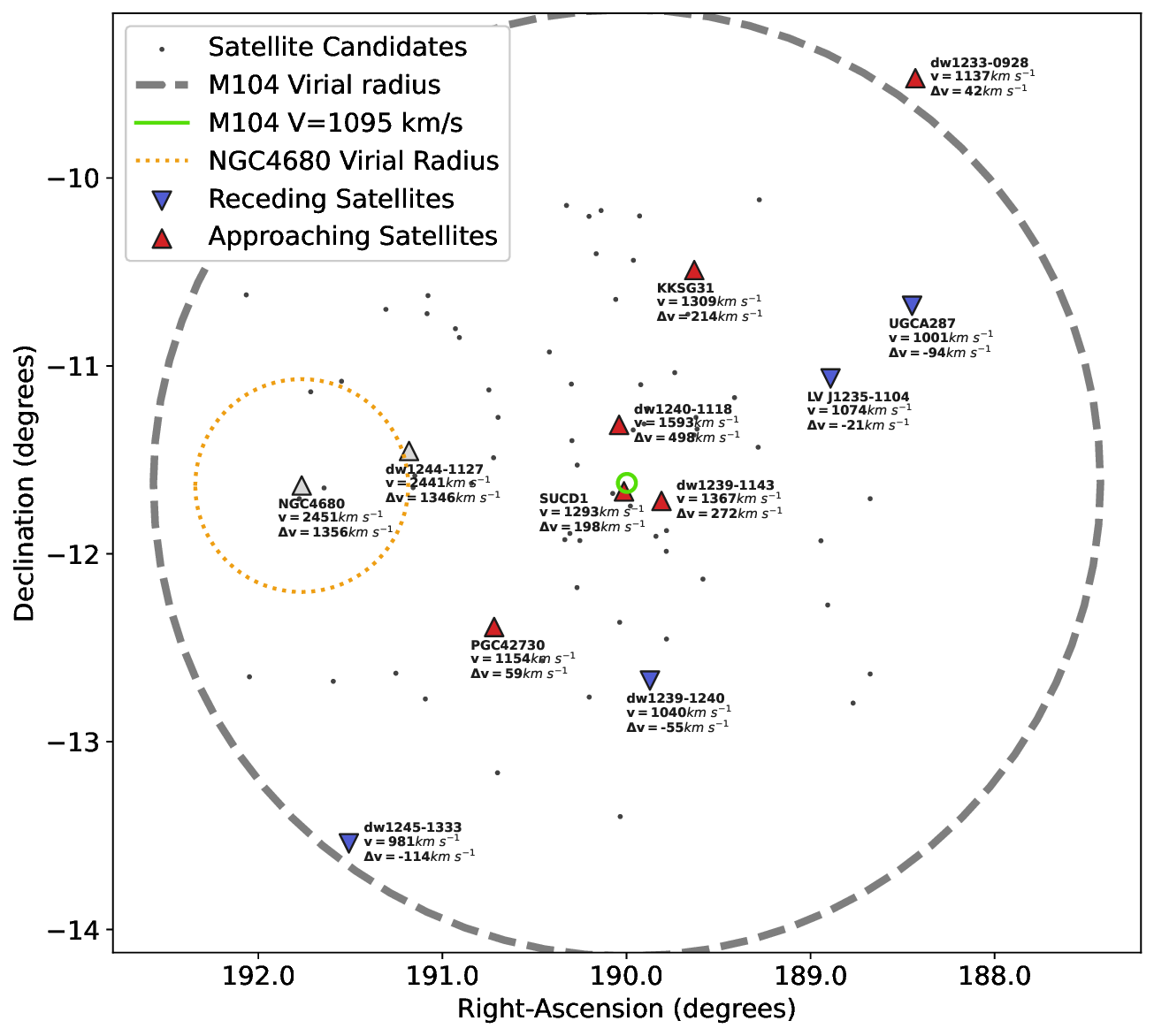}
	\caption{Galaxy distribution in the M104 group environment with measured satellite galaxy velocities. The nine satellites have a mean heliocentric velocity of $1156\,\mathrm{km\,s}^{-1}$ and a velocity dispersion of $221\,\mathrm{km\,s^{-1}}$. Velocities relative to M104 ($1094\,\mathrm{km\,s^{-1}}$) are also given. Symbol colour and the direction of the triangular symbols reflects the relative velocity. As noted in \citet{Crosby_2023_b}, the M104 environment is well defined in velocity space with a significant velocity gap of over $1000\,\mathrm{km\,s}^{-1}$ to the next galaxy grouping in the background that includes NGC4680 ($2451\,\mathrm{km\,s^{-1}}$). The dwarf dw1244-1127 ($2426\,\mathrm{km\,s}^{-1}$) is most likely a companion of NGC4680 as judged from its small relative velocity difference of $25\,\mathrm{km\,s}^{-1}$ and its proximity to NGC4680's virial radius.}
	\label{fig:M104_satellite_map}
\end{figure*}

However, this case only reveals the presence of a satellite plane when such a plane is largely face-on, where the satellite motion would be mostly in its on-sky proper motions relative to the observer and less of their true motion is detected in LOS velocities, thus the LOS velocity distribution appears to colder than less coherently rotating systems. In the case where it is edge-on, we can detect a satellite plane by measuring the fraction of satellites co-rotating in a disk around M104, and compare that to expectations from a non-planar system of satellites. Based on visual inspection of Figure \ref{fig:M104_satellite_map}, there appears to be no preferential co-rotating axis yet, unlike other systems such as NGC4485 / NGC4490 \citep{Karachentsev2024}. But more likely, if a plane is present that can take any orientation, it will be somewhere between face-on and edge-on, and a number of other key parameters, such as the number of satellites in the plane and the plane thickness for example, could also vary significantly based on the diversity of satellite planes already known \citep{Ibata_2013,Conn2013,Mueller2018,Pawlowski_2019}. Ultimately, this expanded analysis would involve comparing the shapes of the satellite LOS velocity distribution, more than just the collective dispersion. We currently lack enough measured recessional velocity from satellites to constrain the uncertainty of such an analysis, so we defer this analysis to later work.

\subsection{Dynamical Mass of M104}\label{dyn_mass}

With the satellite velocity dispersion now measured, we can estimate the dynamical and baryonic mass of the M104 group. We can use the dynamical mass to compare to theoretical mass estimates from \cite{Crosby_2023_b}. From the virial theorem we can use the velocity dispersion to calculate the total mass enclosed within a radius:
\begin{equation}
	M_{tot}=\frac{5R_{max}v_{\sigma}^{2}}{G}
\end{equation}
where $v_{\sigma}^{2}$ is the velocity dispersion as calculated above and $R_{max}$ is the maximum radial extent of the satellite galaxies. For this value, we use the maximum projected radial extent of the satellites multiplied by $\sqrt{2}$, which is $\sim595\,\mathrm{kpc}$. From this, we determine $M_{tot}=(12.4\pm6.5)\times10^{12}M_{\odot}$, which is slightly higher than our theoretical mass estimate from \cite{Crosby_2023_b} of $7.8 \times 10^{12}\,M_{\odot}$, calculated from the stellar mass of M104 and the average stellar mass to total halo mass ratio measured from the TNG100 simulation for similar M104-analogues as described in section \ref{satellite_plane}. This is in line with a similar dynamical mass measurement of $M_{tot}=15.5\pm4.9\times10^{12}M_{\odot}$ from \cite{Karachentsev2020_b} which considered a larger number of dwarf galaxies up to $\sim1\,\mathrm{Mpc}$ away from M104.

The stellar mass of M104 is $M_{*}=1.8\times10^{11}$~M$_{\odot}$ \citep{Mu_oz_Mateos_2015} and likely dominates $>99\%$ of all stellar mass within the virial radius given its brightest satellite PGC42730 has a stellar mass of only $10^{9}\,M_{\odot}$. The stellar masses of the next brightest satellites, among those quoted in Table \ref{tab:IFUMResults}, are even less significant. The baryonic mass fraction of the universe as measured by \cite{Planck2020} is expected to be $\Omega_b/\Omega_m\sim0.16$ and we expect to find a similar ratio within the M104 system. This indicates that the total mass of baryons within M104's virial radius should be $M_{bary}=2.0\pm1.0\times10^{12}M_{\odot}$, based on the dynamical mass measurement above. The stellar mass of M104 is known and the baryonic mass of its satellites is negligible, so there should be a gas mass of $M_{gas}=\sim1.8\times10^{12}M_{\odot}$ within the virial radius. This indicates the gas to stellar mass ratio should be $M_{gas}/M_{*}\sim10$ in order to satisfy the baryonic to dark matter mass ratio, but it is very unlikely that a galaxy such as M104 possess that much gas. A similar class of $L_*$ galaxy Cen A possesses at most $M_{gas}/M_{*}\sim0.2$ \citep{Muller2022} and larger statistical analyses of $L_*$ galaxies suggest that for a galaxy of M104's stellar mass, then this ratio is at most $M_{gas}/M_{*}\sim0.3$ \citep{Warren2007,Parkash2018}. This suggests that M104's baryonic mass to total mass ratio is much smaller than expected, by a factor of $\sim 10$. \cite{Karachentsev2020_b} noted a similar phenomenon with the stellar mass to total mass ratio, which was lower by a factor of $\sim 10$ for that analysis as well. Curiously, our TNG100 sample of simulated M104 analogues have a mean $M_{gas}/M_{*}$ ratio of $M_{gas}/M_{*}=3.8\pm2.0$ for a mean stellar mass of $M_{*}=(0.9\pm0.4)\times10^{11}$~M$_{\odot}$, which is consistently higher than that expected from observations \citep{Warren2007,Parkash2018,Muller2022}. This however still falls short of the expected ratio from observations.

\subsection{Mass Metallicity Relation}

\begin{figure}
	\centering
	\includegraphics[draft=false,width=8cm]{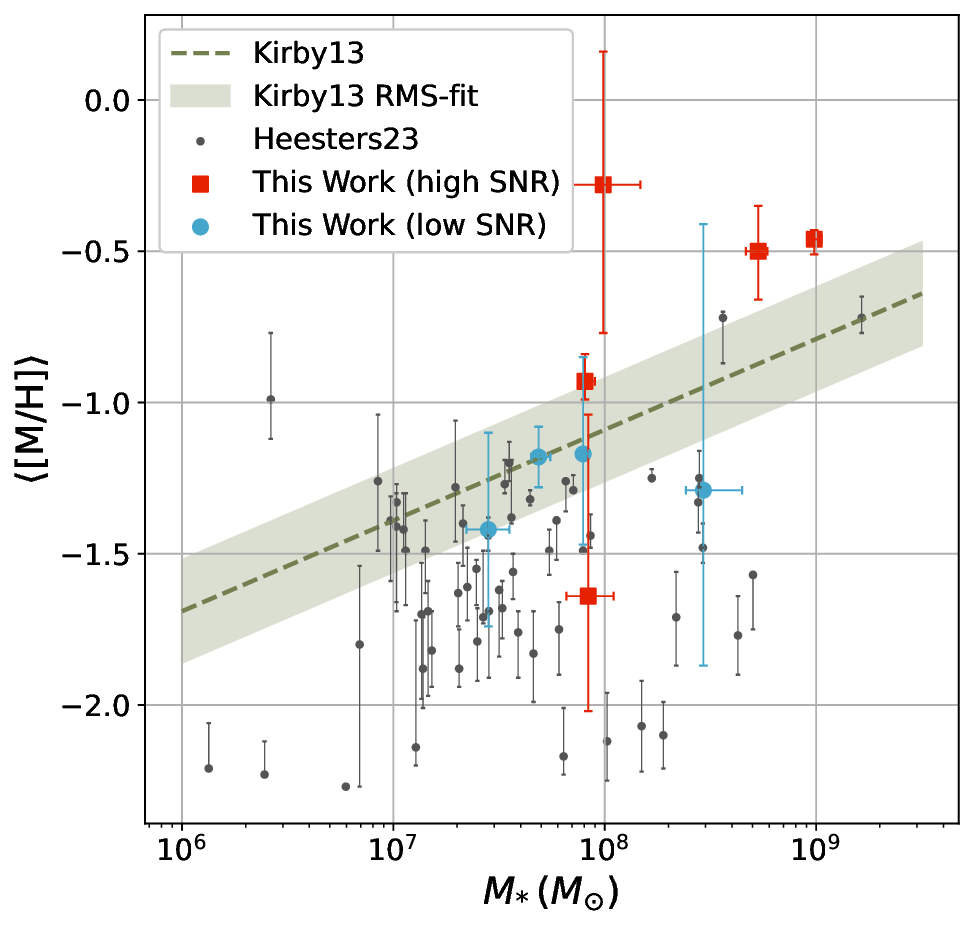}
	\caption{The mass-metallicity relation of the satellites of M104 as presented here, split between those with high SNR ($>20$, shown with squares) and low SNR ($\geq20$, shown with large circles), the sample of dE galaxies from \citet{Heesters2023} (small circles) and the root mean square fit of the mass-metallicity relation for Milky Way dwarf satellites from \citet{Kirby_2013}.}
	\label{fig:M104_MZ}
\end{figure}

We show the mass-metallicity relation for the newly observed M104 satellite galaxies in Figure \ref{fig:M104_MZ}. We compare the M104 satellites to the dwarf ellipticals observed using MUSE in \citet{Heesters2023}. We also plot the root mean square fit of the mass-metallicity relation for the Milky Way dwarfs from \citet{Kirby_2013}. \citet{Heesters2023} found that their sample of early-type dwarf galaxies were mostly more metal poor than the relation found from the Milky Way dwarfs. Here we find that within 1$\sigma$ uncertainty, our M104 dwarfs are consistent with the Milky Way trend. We note that some of spectra have low SNR (SNR<20), which may produce biased stellar ages and metallicities \citep{Woo_2024} and thus we have shown these data points separately in Figure \ref{fig:M104_MZ}. This suggests that the chemical evolution of the M104 environment followed similar pathway as the Milky Way environment, despite the key differences, such as the abundance of dwarf satellites and the presence of nucleated dE galaxies.

\section{Summary and Conclusion}

In this paper we have presented details spectroscopy of ten dwarf galaxies in the vicinity of M104, nine of which we can conclude are satellites of M104 based on recessional velocities alone. Of these nine dwarf galaxies, one is a BCD, two are dwarf irregular, five are nucleated dwarf elliptical galaxies and one is a dwarf elliptical galaxy with no nucleus. The remaining background galaxy is a dwarf irregular.

These galaxies collectively span stellar masses $2\times10^{7}\,M_{\odot}$ to $1\times10^{9}\,M_{\odot}$. For the dwarf ellipticals, the mean mass weighted ages of these dwarf galaxies span the range $7.0-11.0$ Gyr with metallicities $-1.7$ to $-0.3$. The dwarf irregulars UGCA287, dw1233-0928 and dw1244-1127 follow similar trends with the exception of the presence of a younger stellar population. These ages and metallicities match expectations of the historical formation scenarios of dwarf elliptical and irregular galaxies, which suggest these galaxies collectively formed early in the universe, around $\sim7-12\,$Gyr ago.

This spectroscopy includes one BCD and candidate Green Pea, LV J1235-1104. This low mass galaxy ($M_{*}=8.4\times10^{7}\,M_{\odot}$) has extreme emission features powered by star formation, the equivalent width of [\ion{O}{III}] $\lambda 5007$\AA$\ $is 314\AA$\ $and we have detected broad components to emission lines with $\sigma\sim250\,$\kms, consistent with the spectral properties of higher redshift starburst dwarf galaxies named Extreme Emission Line Galaxies (EELGs) or Green Peas (GPs). It is forming stars at a rate of $0.042\,M_\odot$\,yr$^{-1}$ as measured from the $H_{\alpha}$ flux, which gives it a specific star formation rate (sSFR) of $5\times 10^{-10}$\,yr$^{-1}$ which however falls short of typical values for EELGs or GPs. This galaxy contains two powerful star forming regions, whose intense radiation field is producing a galactic outflow of hot ionised gas filling the circumgalactic medium about this galaxy, and optical emission features are visible in our spectra up to $\sim1\,\mathrm{kpc}$ away from the star formation regions. The similarity of this galaxies unique emission features to those of GPs makes this galaxy a candidate Ly$\alpha$/LyC photon leaking galaxy, like higher redshift starburst galaxies. This class of galaxy in the Local Volume is exceedingly rare, and are not usually satellite galaxies, making this satellite particularly unique. As with most low-mass starburst dwarf galaxies it is not immediately obvious what triggered this galaxy's starburst, but given that environmental interactions are likely to rapidly strip a low mass galaxy of this kind of its star forming gas, it is likely a recent addition to the M104 system of satellites.

We consider the bulk recessional velocities of these galaxies in the context of satellite planes. By comparing the dispersion in of the LOS velocities sample of M104 satellites producing by Bootstrapping to that of simulated M104 analogues in the Illustris TNG100-1 simulation, we find no significant difference between the two, indicating that a satellite plane that contends with $\Lambda$CDM simulations of the kind around the Milky Way, M31, or Cen A is not currently detected around M104. However, with only 10 out of 70 possible satellite candidates around M104 within the virial radius confirmed with spectroscopy, ongoing observations and re-testing of this hypothesis will be required.

Finally, we use our new recessional velocities to measure the dynamical mass of M104 within the virial radius. We measure a mass of $M_{\text{tot}}=(12.4\pm6.5)\times10^{12}\,M_{\odot}$ and thus a revised virial radius of $R_{vir}=491\pm86\,\mathrm{kpc}$. Considering M104's stellar mass ($M_{*,M104}=1.8\times 10^{11}M_{\odot}$), the likely gas content for an M104-like galaxy ($M_{gas}\sim0.25\,M_*$), indicates that $>$90\% of its expected baryons are missing, given expected baryon mass fractions from cosmological observations. This matches similar findings from prior measurements of M104's dynamical mass, and of Cen A \citep{Mueller2021}, a morphologically similar galaxy to M104.

\section*{Acknowledgements}


Marcel S. Pawlowski acknowledges funding via a Leibniz-Junior Research Group (project number J94/2020).

Ivanne Escala acknowledges generous support from a Carnegie-Princeton Fellowship through Princeton University.

\section*{Data Availability}
The data underlying this article will be shared on reasonable request to the corresponding author.



\bibliographystyle{mnras}
\bibliography{spec_satellites.bib} 


\clearpage 
\appendix

\bsp	
\label{lastpage}
\end{document}